\documentstyle[preprint,aps]{revtex}
\begin{document}
\draft
%\preprint{MAD/TH/92-7}
\title{%
Two-loop unitarity constraints on
the Higgs boson coupling}
\author{ Loyal
Durand\cite{ld}, Peter N. Maher\cite{pm}, and Kurt Riesselmann\cite{kr}}
\address{%
Department of Physics, University of
Wisconsin--Madison\\
Madison, Wisconsin 53706}
\maketitle
\begin{abstract}
We use the results of Maher {\em et al.\/} (preceding paper) to
construct the matrix of $j=0$ partial-wave two-body and
$2\rightarrow3$ scattering amplitudes for the scattering of
longitudinally polarized gauge bosons $W_L^\pm$, $Z_L$ and Higgs
bosons $H$ correct to two loops in the high-energy, heavy-Higgs
limit $\sqrt{s}\gg M_H\gg M_W$. We show explicitly that the energy
dependence of the $2\rightarrow2$ amplitudes can be completely
absorbed into a running quartic Higgs coupling $\lambda_s=
\lambda_s(s,M_H^2)$ and factors which involve small anomalous
dimensions and remain near unity. After diagonalizing the matrix of
partial-wave amplitudes, we use an Argand-diagram analysis to show
that the elastic scattering amplitudes are approximately unitary and
weakly interacting for $\lambda_s\alt2.3$, but that three-loop
corrections are necessary to restore unitarity for larger values of
$\lambda_s$. That is, the interactions in the Higgs sector of the
standard model are effectively strong with respect to the
perturbative expansion for $\lambda_s\agt2.3$. The bound
$\lambda_s\alt2.3$ for a weakly interacting theory translates to a
physical Higgs mass $M_H\alt380$ GeV if the bound is to hold for
energies up to a few TeV, or $M_H\leq155$ GeV in perturbatively
unified theories with mass scales of order $10^{16}$ GeV.
\end{abstract}

\narrowtext
\section{INTRODUCTION}

In the absence of direct observations of the Higgs boson of the
standard model of electroweak interactions, its mass $M_H$ and
quartic coupling $\lambda=M_H^2/2v^2$ remain unknown parameters in
the theory related through the known value of the vacuum expectation
value $v=\left( \sqrt{2}G_F \right)^{-1/2} =246$ GeV. It is
therefore of interest to search for bounds on $M_H$, and to consider
the possibility that $\lambda$ is large enough that the theory
becomes strongly interacting in the Higgs sector\cite{lqt,velt,cg}.
The only strict upper bound on $M_H$, $M_H\alt 650$--800 GeV,
follows from the so-called triviality bound in theories with
elementary scalar fields \cite{dn} as implemented for the standard
model in lattice calculations \cite{hasen}. A detailed analysis
\cite{djm} of two-body scattering in the Higgs sector of the theory
shows that strong-interaction effects would be essentially invisible
for $\sqrt{s}\ll M_H$ unless $M_H$ is very large, $M_H>$ 4--5 TeV.
Such a mass would violate the triviality bound, but could be allowed
if the standard model is just a low-energy remnant of a more
complete theory in which, for example, the Higgs boson is composite.

The suppression of the low-energy scattering is a general
consequence of chiral symmetry and the known constraints on
electroweak symmetry breaking \cite{cg,cgg}.
In contrast,
high-energy scattering is not suppressed. The
triviality limit
on $\lambda$, $\lambda\alt3.5$,
is large enough that the standard model {\em can} be strongly
interacting
with the effects visible
at energies $\sqrt{s}\gg M_H$ as judged by a detailed
analysis of the scattering amplitudes in the Higgs sector
calculated to one loop \cite{djl1,djl2}. That analysis indicated
that strong coupling sets in  substantially below the tree-level bound
$\lambda\approx 8\pi/3=8.38$ (or $M_H=1.007$ TeV) derived some
time ago by Dicus and Mathur \cite{dm} and Lee, Quigg, and Thacker
\cite{lqt}. The question naturally arises as to whether the bound
would be further strengthened---or substantially weakened---by
extending the analysis to two loops. The question is especially
pertinent because the analysis in \cite{djl1} and \cite{djl2}
used renormalization-group-improved scattering amplitudes
expressed as series in the running coupling $\lambda _s$. An effectively
strongly interacting theory appears for $\lambda _s\agt 2$--2.5 (see also
\cite{mvw,dw1,pass1,dw2,pass2,vay,lend} for less restrictive analyses). The
limits are less obvious in the expanded form of the amplitudes restricted
to terms of orders $\lambda $ and $\lambda ^2$ with energy-dependent
coefficients.

Our objective here is to extend the previous analyses to two loops
using the results for the two-body scattering amplitudes for
longitudinally polarized gauge bosons $W_L^\pm$, $Z_L$ and Higgs
bosons $H$ obtained by Maher {\em et al.\/} \cite{mdr} in the
preceding paper. We explore the range of validity of the two-loop
results with increasing $\lambda $ by making an Argand-diagram
analysis of the diagonalized $j=0$ partial-wave scattering
amplitudes for two-body scattering in the neutral channels
$W_L^+W_L^-,Z_LZ_L,HH$, and $Z_LH$. These amplitudes must lie on or
inside the usual unitarity circle, and we may use major departures
from the circle as evidence of a failure of the series to converge
at order $\lambda ^3$. Quantitative measures of convergence indicate
that the renormalization-group-improved perturbation series
converges very slowly, if at all, for running couplings $\lambda
_s\agt2.3$. The standard model becomes in effect ``electrostrong" in
the Higgs sector for larger values of $\lambda _s$, even though the
perturbative amplitudes are still quite small.

\section{THE SCATTERING AMPLITUDES}

\subsection{Background}

It is well known \cite{lqt} that scattering amplitudes for processes
involving longitudinally polarized gauge bosons $W^\pm_L,Z_L$ are
enhanced in the limit $M_H\gg M_W$ by powers of $M_H^2/M_W^2$ relative to the
corresponding amplitudes for transversely polarized gauge bosons. It is
therefore natural to study the scattering of $W_L^\pm,Z_L$ and Higgs
bosons when seeking an upper bound on $M_H$ or, equivalently, on the
quartic Higgs coupling $\lambda =M_H^2/2v^2$. The necessary calculations
can be greatly simplified in the limit of interest, $M_H\gg M_W$, by use
of the Goldstone boson equivalence theorem
\cite{lqt,cg,corn,bs,hvelt,he}. This theorem states that the scattering
amplitudes for processes which involve $n$ longitudinally polarized gauge
bosons and any number of other external particles, are related to the
corresponding scattering amplitudes for the scalar bosons $w^\pm,z$ to
which $W_L^\pm,Z_L$ reduce in the limit of vanishing gauge couplings
$g,g'$ by
%
% equation
%
\begin{equation}
\label{ii1}
	T(W_L^\pm,Z_L,H,\ldots)=(iC)^nT(w^\pm,z,H,\ldots).
\end{equation}
The corrections are of orders $M_W/\sqrt{s}$ and $g^2,{g'}^2$. In the
renormalization scheme used in \cite{mdr}, $C=1+{\rm O}(g^2,{g'}^2)$.

Maher {\em et al\/}.\ \cite{mdr} used the equivalence theorem to
calculate the complete matrix of two-body scattering amplitudes for the
neutral channels $w^+w^-,zz,HH,zH$ in the high-energy, heavy-Higgs limit
$\sqrt{s}\gg M_H\gg M_W$. (The charged channels provide no extra
information, as we will see.) The high-energy limit $\sqrt{s}\gg M_H$,
while not of immediate experimental interest, introduces the further
simplification that only the dimension-four quartic couplings contribute
to the two-body scattering graphs to leading order in $s$. The
dimension-three couplings contribute to the final results only through
the renormalization constants, which contain contributions from the
low-energy region $\sqrt{s}\alt M_H$.

Maher {\em et al.\/} \cite{mdr} give the renormalized matrix $\cal M$ of
Feynmann transition amplitudes as
%
% equation
%
\begin{equation}
\label{ii2}
	{\cal M}={\bf Z}{\cal F}_R{\bf Z},
\end{equation}
where ${\cal F}_R$ is a matrix of finite, partially renormalized
amplitudes $A_R(s,t,u)$, and ${\bf Z}$ is a finite diagonal matrix of
ratios of renormalization constants. With the two-body channels taken in
the order $w^+w^-,zz,HH$, and $zH$, ${\cal F}_R$ has the form
%
% equation
%
\begin{equation}
\label{ii3}
	{\cal F}_R=\left(
	\begin{array}{cccc}
		A_R(s)+A_R(t) & A_R(s) & A_R(s) & 0\\
		A_R(s) & A_R(s)+A_R(t)+A_R(u) & A_R(s) & 0\\
		A_R(s) & A_R(s) & A_R(s)+A_R(t)+A_R(u) & 0\\
		0 & 0 & 0 & A_R(t)
	\end{array}
	 \right),
\end{equation}
while
%
% equation
%
\begin{equation}
\label{ii4}
	{\bf Z}={\rm diag}\left( 1,1,Z_H/Z_w,\left( Z_H/Z_W \right)^{1/2}
	\right).
\end{equation}
The renormalization constants $Z_w$ and $Z_H$ are given in the
Appendix in Eqs.\ (A4) and (A5).

We have indicated only the first variable in $A_R(s,t,u)$ in Eq.\
(3) since this
function is symmetric under an interchange of the last two variables.
Specifically \cite{mdr},
%
% eqnarray
%
\begin{eqnarray}
A_R(s,t,u)  &=& -2\lambda +\frac{\lambda ^2}{16\pi ^2}\left( -16\ln(-\hat
s)-4\ln(-\hat t)-4\ln(-\hat u)+2+6\sqrt{3}\pi  \right)\nonumber \\
&& + \frac{\lambda ^3}{(16\pi ^2)^2}\left(
\begin{array}{l}
	-192\ln^2(-\hat s)+176\ln(-\hat s)+96\sqrt{3}\pi \ln(-\hat s)\\
	-48\ln^2(-\hat t)+80\ln(-\hat t)+24\sqrt{3}\pi \ln(-\hat t)\\
	-48\ln^2(-\hat u)+80\ln(-\hat u)+24\sqrt{3}\pi \ln(-\hat u)\\
	+60\sqrt{5}\ln\frac{\sqrt{5}+1}{2}-456\sqrt{3}{\bf C}+138\sqrt{3}\pi \\
	+240\ln^2\frac{\sqrt{5}+1}{2}-\frac{3968}{9}\ln2-180\pi ^2\\
	-72K_2-\frac{794}{3}\zeta (2)+180\zeta (3)\\
	-324K_5+24K_3+\frac{3388}{27}-162\zeta (2)\ln2
\end{array}
 \right)\label{ii5}
\end{eqnarray}
where $\hat s=s/M_H^2,\ \hat t=t/M_H^2$, and $\hat u=u/M_H^2$. The
phases are defined so that
$-\hat s=e^{-i\pi }\hat s$,
while the variables
$-\hat t$ and $-\hat u$ are real and
positive in the physical scattering region. $\zeta (n)$ is the
Riemann zeta function, ${\bf C}=1.01494\ldots$ is the value of the
Clausen function at argument $\pi /3$, and the $K$'s are certain
constants evaluated by numerical integration \cite{mdr},
$K_2=-0.86518,\ K_3=-0.10666,\ K_5=0.92363$.

\subsection{The running coupling and anomalous dimensions}

A renormalization group analysis of the scattering amplitudes indicates
that their entire
energy dependence can be subsumed in the limit of interest
($s\gg M_H^2$) into a running coupling $\lambda _s=\lambda _s(s,M_H)$ and
factors involving the anomalous dimensions $\gamma _w,\gamma _z$ and
$\gamma _H$ associated with the $w^\pm,z$ and Higgs bosons. The SO(3)
symmetry of the theory in the $w^\pm,z$ sector gives $\gamma _z=\gamma
_w$ \cite{mdr}. If all momenta are scaled by a factor
$\sigma $ so that $s,t,u\rightarrow\sigma ^2s,\sigma ^2t,\sigma ^2u$, the
scaled and original scattering amplitudes are related by \cite{chli}
%
% equation
%
\begin{equation}
\label{ii6}
	{\cal M}\left( \left\{ \sigma p_i \right\},\lambda ,M_H
	\right)=\mbox{\boldmath$\Gamma $}{\cal M}\left( \left\{ p_i
	\right\},\lambda _s(\sigma ^2s,M_H),M_H \right)\mbox{\boldmath$\Gamma $}.
\end{equation}
Here $\mbox{\boldmath$\Gamma $}$ is a diagonal matrix,
%
% equation
%
\begin{equation}
\label{ii7}
	\mbox{\boldmath$\Gamma $}={\rm diag}(\Gamma _w^2,\Gamma _w^2,\Gamma
	_H^2,\Gamma _w\Gamma _H),
\end{equation}
with
%
% equation
%
\begin{equation}
\label{ii8}
	\Gamma _i={\rm exp}\left( -\int_{\lambda _s(1)}^{\lambda _s(\sigma
	)}\frac{\gamma _i(\lambda )}{\beta (\lambda )}d\lambda  \right),\qquad
	i=w,H,
\end{equation}
and $\lambda _s(\sigma )\equiv \lambda _s(\sigma ^2s,M_H)$. The evolution
of $\lambda _s$ is determined by the equation
%
% equation
%
\begin{equation}
\label{ii9}
	\ln\sigma =\int_{\lambda _s(1)}^{\lambda _s(\sigma )}\frac{d\lambda
	}{\beta (\lambda )}.
\end{equation}

It was demonstrated explicitly in [8,9] that the entire energy dependence
of the scattering amplitudes calculated to one loop could be absorbed in
a running coupling $\lambda _s=\lambda _s(s,M_H^2)$. To that order,
$\beta ^{(1)}(\lambda )=24\lambda ^2/16\pi ^2$ and the $\gamma $'s
vanish. The running coupling $\lambda _s$ was defined in those
references, following the definition given by Sirlin and Zucchini
\cite{sz} for the complete gauge theory, to incorporate some naturally
occurring constants as well as the usual leading-logarithmic dependence
on $s$,
%
% equation
%
\begin{equation}
\label{ii10}
	\lambda _s(s,M_H)=\lambda \left[ 1-\frac{\lambda }{16\pi ^2}\left(
	24\ln\frac{\sqrt{s}}{M_H}+25-3\sqrt{3}\pi  \right) \right]^{-1}.
\end{equation}
The constant $\lambda $ is then related to the muon decay constant and
the physical mass of the Higgs boson by [25]
%
% equation
%
\begin{equation}
\label{ii11}
	\lambda =M_H^2/2v^2=G_\mu M_H^2/\sqrt{2}.
\end{equation}

The calculation of $\gamma _w,\gamma _H$, and $\beta$ at two loops
involves some unusual features because of our on-mass-shell
renormalization conventions, and is considered in Appendix A. The results
are:
%
% eqnarray
%
\begin{eqnarray}
\label{ii12}
	\gamma _w  &=& -6\left( \frac{\lambda }{16\pi ^2} \right)^2+{\rm
	O}(\lambda ^3),\\
	\gamma _H  &=& (150-24\pi \sqrt{3})\left( \frac{\lambda }{16\pi ^2}
	\right)^2+{\rm O}(\lambda ^3),\label{ii13}\\
	\beta   &=& 24\frac{\lambda ^2}{16\pi ^2}\left( 1-13\frac{\lambda
	}{16\pi ^2} \right)+{\rm O}(\lambda ^3).\label{ii14}
\end{eqnarray}

Integration of Eq.\ (\ref{ii9}) gives the running coupling $\lambda _s$,
%
% equation
%
\begin{equation}
\label{ii15}
	\frac{1}{\lambda _s(\sigma )}=\frac{1}{\lambda _s(1)}-(\beta _0+\beta
	_1\lambda )\ln\sigma +{\rm O}(\lambda ^2),
\end{equation}
or with $\sigma $ chosen as the ratio of $\sqrt{s}$ to the energy scale
at which $\lambda _s=\lambda $,
%
% equation
%
\begin{equation}
\label{ii16}
	\lambda _s(s,M_H)=\lambda \left[ 1-\frac{\lambda }{16\pi ^2}\left(
	1-\frac{13\lambda }{16\pi ^2}\right)
	\left(
	24\ln\frac{\sqrt{s}}{M_H}+25-3\sqrt{3}\pi  \right) \right]^{-1}.
\end{equation}
It is interesting to note that the Landau pole in $\lambda _s$ apparently
disappears for $\lambda >16\pi ^2/13\approx12$ or $M_H>1214$ GeV when
the second term in the denominator changes sign. However, as we will see
later, such values of $\lambda $ are far outside the range in which a
two-loop calculation is reliable.

With the results above, the factors $\Gamma _i$ in Eq.\ (\ref{ii8}) can
be written as
%
% equation
%
\begin{equation}
\label{ii17}
	\Gamma _i={\rm exp}\left( -\,\frac{\gamma _{0,i}}{\beta _0}
	\,\frac{\lambda
	_s(\sigma )-\lambda _s(1)}{16\pi ^2} \right),
\end{equation}
where $\gamma _{0,1}$ and $\beta _0$ are the coefficients of $\lambda ^2$
in Eqs.\ (\ref{ii12})--(\ref{ii14}). With the use of the one-loop running
coupling in Eq.\ (\ref{ii10}), this becomes
%
% equation
%
\begin{equation}
\label{ii18}
	\Gamma _i={\rm exp}\left( -\gamma _{0,1}\left( \frac{\lambda _s(\sigma
	)}{16\pi ^2} \right)^2\ln\sigma +{\rm O}(\lambda _s^3) \right).
\end{equation}
The $\lambda _s$ which appears in this equation is to be interpreted as
the running coupling in accord with the renormalization group analysis.
The scale $\sigma $ will be defined as above so that
%
% equation
%
\begin{equation}
\label{ii19}
	\beta _0\ln\sigma =24\ln\frac{\sqrt{s}}{M_H}+25-3\sqrt{3}\pi ,
\end{equation}
or
%
% equation
%
\begin{equation}
\label{ii20}
	\sigma =\frac{\sqrt{s}}{M_H}{\rm exp}\left( \frac{25-3\sqrt{3}\pi
	}{24} \right)\approx1.43\frac{\sqrt{s}}{M_H}.
\end{equation}
With these conventions understood,
%
% equation
%
\begin{equation}
\label{ii21}
	\Gamma _i=\sigma ^{-\gamma _i(\lambda _s)}.
\end{equation}

It is straightforward to check that all the energy dependence of the
scattering amplitudes $\cal M$ can indeed be absorbed in the running
coupling and the anomalous-dimension factors $\gamma _i$ as indicated in
Eq.\ (\ref{ii6}) \cite{algebra}. Specifically,
%
% equation
%
\begin{equation}
\label{ii22}
	{\cal M}(s,\lambda )=\mbox{\boldmath$\Gamma $}(s,\lambda
	_s)\widehat{\cal M}(\cos\theta ,\lambda _s(s,M_H))\mbox{\boldmath$\Gamma
	$}(s,\lambda _s)
\end{equation}
where $\widehat{\cal M}$ depends on $s$ only through the running
coupling, and is otherwise a function only of the scattering angle. The
anomalous dimensions $\gamma _w$ and $\gamma _H$ are quite small
numerically for $\lambda _s$ in the range of interest,
%
% eqnarray
%
\begin{eqnarray}
\label{ii23}
	\gamma _w  &=& -0.000240\lambda _s^2,\\
	\gamma _H &=& 0.000778\lambda _s^2.
\end{eqnarray}
It will therefore be a good approximation in parts of the later analysis
to replace the diagonal matrix $\mbox{\boldmath$\Gamma $}$
in Eq.\ (\ref{ii22}) by $\bf 1$.

\subsection{Partial-wave $2\rightarrow2$ amplitudes for $j=0$}

Our later analysis of (apparent) violations of unitarity in perturbation
theory will be based on the known properties of partial-wave scattering
amplitudes. The matrix ${\bf a}_j^{2\rightarrow2}$ of $2\rightarrow2$
partial-wave scattering amplitudes for angular momentum $j$ is related to
the matrix $\cal M$ of Feynman amplitudes for the various channels by
\cite{djl2,ldjl}
%
% equation
%
\begin{equation}
\label{ii25}
	{\bf a}_j^{2\rightarrow2}(s)={\bf N}{\cal A}_j{\bf N}
\end{equation}
where ${\cal A}_j$ is the properly normalized partial-wave projection of
$\cal M$,
%
% equation
%
\begin{equation}
\label{ii26}
	{\cal A}_j(s)=\frac{1}{32\pi} \left( \frac{4p_ip_f}{s}
	\right)^{1/2}\int_{-1}^{1}d\cos\theta {\cal M}(s,\cos\theta)
	P_j(\cos\theta ).
\end{equation}
The momentum-dependent prefactor approaches unity for $\sqrt{s}\gg M_H$.
The matrix $\bf N$ incorporates the symmetry factors which must be
inserted for each pair of identical particles in the initial and/or final
state \cite{djl2,ldjl}. It is given for channel labels $w^+w^-,zz,HH,zH$ by
%
% equation
%
\begin{equation}
\label{ii27}
	{\bf N}={\rm diag}(1,\ 1\sqrt{2},\ 1/\sqrt{2},\ 1).
\end{equation}
corresponding to two-body states
$|w^+w^-\rangle,\frac{1}{\sqrt{2}}|zz\rangle,\frac{1}{\sqrt{2}}|HH\rangle$,
and $|zH\rangle$ normalized over the entire solid angle. The matrix
${\bf a}_j^{2\rightarrow2}$ is related to the $S$ matrix by
%
% equation
%
\begin{equation}
\label{ii28}
	{\bf a}_j^{2\rightarrow2}=({\bf S}_j^{2\rightarrow2}-{\bf 1})/2i.
\end{equation}

We will deal only with the $j=0$ partial-wave amplitudes. The scattering
amplitudes for $j>0$ are quite small, and do not give useful unitarity
constraints. The $j=0$ amplitudes are easily calculated, but the results
are too lengthy to record in detail. However, to illustrate their
character, we give the result for the diagonal $zH$ channel:
%
% equation
%
\begin{equation}
\label{ii29}
	a_0^{zH}(s)=\sigma ^{-2\gamma _w-2\gamma _H}\hat a_0^{zH}(s),
\end{equation}
where $\hat a_0^{zH}$ is a cubic polynomial in $\lambda _s$,
%
% eqnarray
%
\begin{eqnarray}
\label{ii30}
	\hat a_0^{zH}(s)  &=& -\frac{2\lambda _s}{16\pi }+\frac{\lambda
	_s}{16\pi }\,\frac{\lambda _s}{16\pi ^2}(46+4\pi \sqrt{3}+4\pi i)\nonumber \\
	&&+\frac{\lambda _s}{16\pi }\left( \frac{\lambda _s}{16\pi ^2} \right)^2
	\left(
	\begin{array}{l}
		-2980+360\zeta (3)-356\zeta (2)-648K_5\\
		-144K_2-640\ln2-324\zeta (2)\ln2\\
		+48\ln^2\frac{\sqrt{5}+1}{2}+672\pi \sqrt{3}-1124\sqrt{3}{\bf C}\\
		+32\pi \sqrt{3}\ln3+144\sqrt{5}\ln\frac{\sqrt{5}+1}{2}\\
		-162\pi ^2-i(228\pi +8\pi ^2\sqrt{3})
	\end{array}
	 \right).
\end{eqnarray}

\subsection{The $2\rightarrow3$ amplitudes and unitarity}

The unitarity of the $S$ matrix provides a nontrivial check on the
calculation of the scattering amplitudes. The relation $\bf S^\dagger
S=1$ reduces for the partial-wave amplitudes to the relation
%
% equation
%
\begin{equation}
\label{ii31}
	{\rm Im}\,{\bf a}_j^{2\rightarrow2}=({\bf
	a}_j^{2\rightarrow2})^\dagger{\bf a}_j^{2\rightarrow2}+\sum_{n>2}\left(
	{\bf a}_j^{2\rightarrow n} \right)^\dagger{\bf a}_j^{2\rightarrow n},
\end{equation}
where the generalized sum in the last term includes an integration over
the $n$-particle phase space. Expanding the scattering amplitudes in
power series in $\lambda $ and equating like powers to order $\lambda ^3$
gives the matrix relations
%
% eqnarray
%
\begin{eqnarray}
\label{ii32}
	{\rm Im}\,{\bf a}_j^{(0)} &=& 0,\nonumber \\
	{\rm Im}\,{\bf a}_j^{(1)} &=& {\bf a}_j^{(0)\dagger}{\bf a}_j^{(0)}\nonumber
\\
	{\rm Im}\,{\bf a}_j^{(2)} &=& {\bf a}_j^{(0)\dagger}{\rm Re}\,{\bf
	a}_j^{(1)}+\left( {\rm Re}\,{\bf a}_j^{(1)} \right)^\dagger{\bf
	a}_j^{(0)}+\sum \left( {\bf a}_{j,\rm tree}^{2\rightarrow3}
	\right)^\dagger{\bf a}_{j,\rm tree}^{2\rightarrow3}.
\end{eqnarray}
The last equation relates the imaginary part of the two-loop amplitude to
the one-loop and tree level amplitudes for the $2\rightarrow2$ processes,
and the tree-level contributions to the inelastic $2\rightarrow3$
processes. The diagrams for the $2\rightarrow3$ processes are shown in
Fig.\ 1.

One would normally expect the contributions of the $2\rightarrow3$
processes to vanish for $s\rightarrow\infty$ since the relevant diagrams
involve dimension-three operators. The graphs in Figs.\ 1a and 1b are
suppressed as expected by the propagator of the exchanged particle
(proportional to $1/t$ or $1/u$) and vanish for $s\rightarrow\infty$ as
$\hat s^{-1}\ln\hat s$ when projected to any fixed $j$. However, the
two-particle ``jet" in the graph in Fig.\ 1c can have a low mass so that
the extra propagator is nearly on shell. As a result, the square of this
graph
is not suppressed in the unitarity sum in Eq.\ (\ref{ii31})
after the integration over the 3-body phase space.
However,
interference terms between jet graphs which differ by the
exchange of a jet particle and the final particle from the 4-point vertex
are still suppressed. The only finite contributions of the
$2\rightarrow3$ processes to the unitarity sum are therefore from terms
which have the topology of a cut scattering {\em Eye\/} graph in the
terminology of \cite{mdr}. The three such graphs for $zH$ scattering are
shown in Fig.\ 2. We will consider this set as an example.

The contribution of any of the diagrams in Fig.\ 2 to the unitarity sum
is given up to vertex and symmetry factors by
%
% equation
%
\begin{eqnarray}
\label{ii33}
	&&\frac{|{\bf p}_a|}{4\sqrt{s}}\int\frac{dP_3}{4\pi }\left(
	\frac{i}{s_{12}-m^2+i\epsilon } \right)^*\frac{i}{s_{12}-m^2+i\epsilon }
	\nonumber \\
	&&\raisebox{-1em}{$\stackrel{\displaystyle
	\longrightarrow}{\scriptstyle s\gg M_H^2}$}\ \frac{1}{(16\pi
	)^2}\,\frac{1}{\pi }\int_{(m_1+m_2)^2}^{\infty}ds_{12}\frac{\Delta
	(s_{12},m_1^2,m_2^2)}{16\pi s_{12}\left[ \left( s_{12}-m^2
	\right)^2+\epsilon ^2 \right]}.
\end{eqnarray}
Here $\Delta $ is the triangle function,
%
% equation
%
\begin{equation}
\label{ii34}
	\Delta (a,b,c)=\left[ (a-b-c)^2-4bc \right]^{1/2},
\end{equation}
and $m_1,m_2,\sqrt{s_{12}}$, and $m$ are, respectively, the masses of the
two particles in the jet, the jet itself, and the intermediate particle.
The kinematic factor $|{\bf p}_a|/4\sqrt{s}$ in Eq.\ (\ref{ii33})
normalizes the $2\rightarrow3$ jet amplitudes (which automatically have
$j=0$ at O$(\lambda ^3)$) to the standard $2\rightarrow2$ partial-wave
amplitudes used above \cite{djl2app}.

The calculations for the graphs in Figs.\ 2a and 2b are straightforward.
Summing over the independent diagrams and taking account of the symmetry
factors for identical particles in the final state we obtain the
value
%
% equation
%
\begin{equation}
\label{ii35}
	\frac{\lambda }{(4\pi
	)^2}\left|a_0^{(0)}(zH)\right|^2=-Z_w^{(1)}\left|a_0^{(0)}(zH)\right|^2
\end{equation}
for the graph in Fig.\ 2a and
%
% equation
%
\begin{equation}
\label{ii36}
	\frac{\lambda }{(4\pi )^2}(-9+2\pi \sqrt{3})\left|a_0^{(0)}(zH)\right|^2
\end{equation}
for the graph in Fig.\ 2b, where $a_0^{(0)}(zH)=-2\lambda /16\pi $ is the
tree-level amplitude for $zH$ scattering.

The graph in Fig.\ 2c is singular for $\epsilon =0$ (that is, for
Im$s_{12}\rightarrow0^+$) since the decays $H\rightarrow w^+w^-$ and
$H\rightarrow zz$ are allowed and the intermediate propagator can be on
mass shell. A direct calculation retaining the $\epsilon $ in the
denominator in Eq.\ (\ref{ii32}) gives
%
% equation
%
\begin{equation}
\label{ii37}
	\frac{3\lambda }{(4\pi )^2}\left( \frac{\pi M_H^2}{\epsilon }-1
	\right)\left|a_0^{(0)}(zH)\right|^2,
\end{equation}
where $\epsilon $ can be identified in the narrow-width approximation for
the $H$ decay as $M_H\Gamma _H$. The
tree-level decay width is $\Gamma _H=3\lambda
M_H/16\pi $, so the first term in Eq.\ (\ref{ii37}) is just equal to the
square of the tree-level $zH$ scattering amplitude. This has already been
included in the calculation in the first term on the right hand side of
Eq.\ (\ref{ii31}) since we treated the Higgs boson as stable, and should
be dropped here---the $H$ always decays. (Alternatively, had we recognized
from the start that the $H$ does not appear as an asymptotic state, we
would have had no tree-level amplitude for $zH$ scattering; it would
appear as here when the $H$ resonance is treated as narrow in
intermediate states.)

With this specification, the combined result for Eqs.\ (\ref{ii36}) and
(\ref{ii37}) reduces to
%
% equation
%
\begin{equation}
\label{ii38}
	\frac{\lambda }{(4\pi )^2}(-12+2\pi
	\sqrt{3})\left|a_o^{(0)}(zH)\right|^2
	=-Z_H^{(1)}\left|a_0^{(0)}(zH)\right|^2,
\end{equation}
where $Z_H^{(1)}$ is the one-loop contribution to $Z_H$.
Upon substituting the results in Eqs.\ (\ref{ii35}) and (\ref{ii38}) into
Eq.\ (\ref{ii32}), we find that the unitarity condition is satisfied in
the channel $zH\rightarrow zH$.

Similar results hold in other channels for the remaining non-suppressed
$2\rightarrow3$ contributions. In each case the relevant graphs have the
toplogy of cut scattering {\em Eye\/} graphs, possibly connecting
different 2-body states, and give results which are just the negative of
the first-order renormalization constant $Z^{(1)}$ for the intermediate
particle in the jet, multiplied by a product of a tree-level amplitudes.
This is not accidental: after the tree-level amplitudes are extracted,
the integrals which remain are closely related to dispersion relations
for the renormalization constants. In particular, the self-energy
functions satisfy once-subtracted dispersion relations
%
% equation
%
\begin{equation}
\label{ii39}
	{\rm Re}\Pi (p^2)-{\rm Re}\Pi (m^2)=\frac{p^2-m^2}{\pi }{\cal
	P}\int_{(m_1+m_2)^2}^{\infty}ds'\frac{{\rm Im}\Pi
	(s'\,)}{(s'-p^2)(s'-m^2)}.
\end{equation}
Dividing by $(p^2-m^2)$, taking the limit $p^2\rightarrow m^2$, and using
the definition of $Z$  in terms of the physical fields and
self-energy functions gives
%
% equation
%
\begin{equation}
\label{ii40}
	Z-1=\frac{1}{\pi }{\cal P}\int_{(m_1+m_2)^2}^{\infty}ds'\frac{{\rm
	Re}\Pi (s')}{(s-m^2)^2}.
\end{equation}

The integral in Eq.\ (\ref{ii33}) is already in the form of a dispersion
relation for $m^2<(m_1+m_2)^2$ since $\epsilon $ can then be set equal to
zero. If $m^2>(m_1+m_2)^2$, the integration contour is trapped between
poles at $m^2\pm i\epsilon $, but by moving the contour outside the
poles, once above and once below and averaging, we pick up a term
proportional to $\epsilon ^{-1}{\rm Im}\Pi (m^2)$ and a principal value
integral. Once the former is deleted, the integrals which remain give the
renormalization constants. In particular, the product of the 3-point
vertex factors in Fig.\ 2a and the factor $\Delta /16\pi s_{12}$ in Eq.\
(\ref{ii33}) is equal to $-{\rm Im}\Pi _w^{(1)}(s_{12}$), where the minus
sign arises from the difference between the factor $(2i\lambda v)^2$
which appears in $\Pi _w^{(1)}$ and the factor $|2i\lambda v|^2$ which
appears in the absolute square of the $2\rightarrow3$ amplitude. The sum
of the corresponding contributions from Figs.\ 2b and 2c gives $-{\rm
Im}\Pi _H^{(1)}$. This analysis clearly generalizes to higher orders.

The final result for the $2\rightarrow3$ contributions to the last of
Eqs.\ (\ref{ii32})---with the $H$ decay term eliminated---is of the form
%
% equation
%
\begin{equation}
\label{ii41}
	{\bf a}_0^{(0)T}\left( -\sum{\bf Z}^{(1)} \right){\bf a}_0^{(0)}
\end{equation}
where $\sum{\bf Z}^{(1)}$ is a diagonal matrix composed of the sums of
the first-order renormalization constants for the final two particles in
the amplitude ${\bf a}_{(0)}$, e.g., $Z_w+Z_H$ for the $zH$ channel. With
this specification, Eqs.\ (\ref{ii32}) are satisfied identically as
matrix equations, a useful and nontrivial check on the calculations.

\section{ANALYSIS AND CONCLUSIONS}

\subsection{Diagonalization of the partial-wave amplitudes}

The general unitarity relation for the $S$ matrix, $\bf S^\dagger S=1$,
reduces in a basis in which the matrix ${\bf a}_j^{2\rightarrow2}$ of
$2\rightarrow2$ partial-wave amplitudes is diagonal to the condition
%
% equation
%
\begin{equation}
\label{iii42}
	\left|a_j^{2\rightarrow2}-\frac{i}{2}\right|^2+\sum_{n>2}\left( {\bf
	a}_j^{2\rightarrow n} \right)^\dagger{\bf a}_j^{2\rightarrow
	n}=\frac{1}{4}.
\end{equation}
This gives the familiar constraint that the exact two-body elastic
scattering amplitudes lie on or inside a circle of radius $\frac{1}{2}$
centered at $\left( 0,\frac{1}{2} \right)$ in the complex plane, a
constraint which we will use to examine the breakdown of low-order
perturbation theory for large values of $\lambda _s$.

The determination of the diagonal $2\rightarrow2$ scattering amplitudes
is straightforward. The matrix $\cal M$ of Feynman amplitudes defined in
Eqs.\ (\ref{ii2})--(\ref{ii4}) would be SO(4) symmetric and could be
diagonalized in an SO(4) basis except for the difference between the
renormalization constraints $Z_w$ and $Z_H$ in the matrix $\bf Z$.
However $\cal M$ and the partial-wave scattering matrices ${\bf
a}_j^{2\rightarrow2}$ retain the exact SO(3) symmetry of the Higgs
Lagrangian \cite{mdr}. We will therefore follow \cite{djl2} and
diagonalize the scattering matrices using states based on the SO(3)
decomposition of SO(4). The sixteen possible two-body combinations of
$w^\pm,z,H$ states break up under SO(4) as $\bf 4\otimes
4=9\oplus6\oplus1$. Because of Bose symmetry, only the symmetric
representations $\bf 9$ and {\bf 1} are allowed for even angular momenta
$j$, and only the antisymmetric representation {\bf 6} for odd $j$. The
SO(4) representations decompose under SO(3) as $\bf
9\rightarrow5\oplus3\oplus1,6\rightarrow3\oplus3'$, and $\bf
1\rightarrow1$. The states we will need for even $j$ are \cite{djl2}
%
% eqnarray
%
\begin{eqnarray}
\label{iii43}
	\bf 9,5  &=& \left\{
	\begin{array}{l}
		w^+w^-\\
		w^+z\\
		\frac{1}{\sqrt{3}}(w^+w^--zz)\\
		w^-z\\
		w^-w^-
	\end{array}
	 \right. ,\quad {\bf 9,3}=\left\{
	 \begin{array}{l}
	 	w^+H\\
	 	zH\\
	 	w^-H
	 \end{array}
	  \right.\nonumber \\
	  \bf 9,1 &=& \frac{1}{\sqrt{24}}(2w^+w^-+zz-3HH),\nonumber \\
	  \bf 1,1 &=& \frac{1}{\sqrt{8}}(2w^+w^-+zz+HH),
\end{eqnarray}
while for $j$ odd the relevant states are
%
% equation
%
\begin{equation}
\label{iii44}
	{\bf 6,3}=\left\{
	\begin{array}{l}
		w^+z\\
		w^+w^-\\
		w^-z
	\end{array}
	 \right. ,\quad {\bf 6,3'}=\left\{
	 \begin{array}{l}
	 	w^+H\\
	 	zH\\
	 	w^-H
	 \end{array}
	  \right.
\end{equation}

Because the SO(3) symmetry is exact, the $\bf (9,5),(9,3)$, and two
$\bf(6,3)$ representations give eigenstates of ${\bf
a}_j^{2\rightarrow2}$ to any order in $\lambda $. However, the two
identity representations of SO(3), $\bf(9,1)$ and $\bf(1,1)$, mix through
the SO(4)-breaking contributions of the $HH$ channel.

The eigenvalues of ${\bf a}_j^{2\rightarrow2}$ for any $j$ are easily
determined by diagonalizing the $4\times 4$ scattering matrix for the
neutral channels. The charged channels add no new information as is
evident from the decompositions above. We will concentrate here on $j=0$,
and will use basis states normalized over the entire solid angle,
specifically the initial states $w^+w^-,zz/\sqrt{2}, HH/\sqrt{2},zH$, and
the SO(3) basis states $\chi _i,i=1,\ldots,4$ defined as
%
% eqnarray
%
\begin{eqnarray}
\label{iii45}
	\chi _1 &=& \chi _{{\bf 1,1}}=\frac{1}{\sqrt{8}}(2w^+w^-+zz+HH),\nonumber \\
	\chi _2 &=& \chi _{{\bf 9,1}}=\frac{1}{\sqrt{24}}(2w^+w^-+zz-3HH),\nonumber \\
	\chi _3 &=& \chi _{{\bf 9,5}}=\frac{1}{\sqrt{3}}(w^+w^--zz),\nonumber \\
	\chi _4 &=& \chi _{{\bf 9,3}}=zH.
\end{eqnarray}
The transformation ${\bf O}^{\rm T}{\bf aO}$ with
%
% equation
%
\begin{equation}
\label{iii46}
	{\bf O}=\left(
	\begin{array}{c@{\quad}c@{\quad}c@{\quad}c}
		\frac{1}{\sqrt{2}} & \frac{1}{\sqrt{6}} & \frac{1}{\sqrt{3}} & 0\\
		\frac{1}{2} & \frac{1}{2\sqrt{3}} & -\sqrt{\frac{2}{3}} & 0\\
		\frac{1}{2} & -\frac{\sqrt{3}}{2} & 0 & 0\\
		0 & 0 & 0 & 1
	\end{array}
	 \right)
\end{equation}
splits off diagonal $\bf 9,5$ and $\bf 9,3$ amplitudes  to any
order in $\lambda $, leaving a $2\times 2$ matrix to be diagonalized in
the $\chi _1,\chi _2$ sector.

We can determine the eigenvalues of the $2\times 2$ matrix rather easily
by transforming to the states which diagonalize the matrix to order
$\lambda ^2$, that is, at one loop \cite{djl2},
%
% eqnarray
%
\begin{eqnarray}
\label{iii47}
	\chi _1' &=& N(\chi _2-\sqrt{3}\Delta \chi _2),\nonumber \\
	\chi _2' &=& N(\chi _2+\sqrt{3}\Delta \chi _1).
\end{eqnarray}
Here $\Delta =\frac{1}{2}\left( Z_H^{(1)}-Z_w^{(1)} \right)$ and
$N=(1+3\Delta )^{-1/2}$. In this basis, the off-diagonal elements of the
$2\times 2$ matrix are of minimum order $\lambda ^3$ and do not
contribute to the eigenvalues to the accuracy of the two-loop
calculation. We conclude, therefore, that the diagonal scattering
amplitudes $a_i$ are given to O($\lambda ^2$) by the diagonal elements of
the transformed matrix,
%
% equation
%
\begin{equation}
\label{iii48}
	a_i=\left( {{\bf O}'}^{\rm T}{\bf O}^{\rm T}{\bf aOO'} \right)_{ii}.
\end{equation}
Here \cite{djl2}
%
% equation
%
\begin{equation}
\label{iii49}
	{\bf O}'=\left(
	\begin{array}{c@{\quad}c@{\quad}c@{\quad}c}
		1-\frac{3}{2}\Delta ^2 & \sqrt{3}\Delta  & 0 & 0\\
		-\sqrt{3}\Delta  & 1-\frac{3}{2}\Delta ^2 & 0 & 0\\
		0 & 0 & 1 & 0\\
		0 & 0 & 0 & 1
	\end{array}
	 \right).
\end{equation}
where we have expanded the normalization factor $N$ to the relevant order.

The diagonal elements $a_1$ and $a_2$ obtained by this construction
appear initially as sums of terms which involve different factors $\Gamma
(s)$ from the anomalous dimensions. Since these factors are close to
unity for the values of $s$ we will consider, it is convenient to expand
the $\Gamma $'s and resum the results to obtain overall effective $\Gamma
$'s for these two scattering eigenstates. The anomalous dimension factors
for $a_3$ and $a_4$ appear just as given below.
The final results are best presented numerically:
%
% eqnarray
%
\begin{eqnarray}
\label{iii50}
	a_1(s) &=& \sigma ^{\gamma _1}\frac{\lambda _s}{16\pi }\left[
	-6+\frac{\lambda _s}{16\pi ^2}(185.6+113.1i) \right.\nonumber \\
	&&\left.-\left( \frac{\lambda _s}{16\pi ^2} \right)^2(8330.9+6892.3i)+{\rm
	O}(\lambda _s^3)\right],\nonumber \\
	a_2(s) &=& \sigma ^{\gamma _2}\frac{\lambda _s}{16\pi }\left[
	-2+\frac{\lambda _s}{16\pi ^2}(65.65+12.57i) \right.\nonumber \\
	&&\left.-\left( \frac{\lambda _s}{16\pi ^2}
	\right)^2(3590.7+839.7i)+{\rm O}(\lambda _s^3)\right],\nonumber \\
	a_3(s) &=& \sigma ^{\gamma _3}\frac{\lambda _s}{16\pi }\left[
	-2+\frac{\lambda _s}{16\pi ^2}(72+12.57i) \right.\nonumber \\
	&&\left.-\left( \frac{\lambda _s}{16\pi ^2}
	\right)^2(4752.5+879.6i)+{\rm O}(\lambda _s^3)\right],\nonumber \\
	a_4(s) &=& \sigma ^{\gamma _4}\frac{\lambda _s}{16\pi }\left[
	-2+\frac{\lambda _s}{16\pi ^2}(67.77+12.57i) \right.\nonumber \\
	&&\left.-\left( \frac{\lambda _s}{16\pi ^2}
	\right)^2(3980.2+853.0i)+{\rm O}(\lambda _s^3)\right],
\end{eqnarray}
where
%
% eqnarray
%
\begin{eqnarray}
\label{iii51}
	\gamma _1 &=& -3\gamma _w-\gamma _H=-24\left( \frac{\lambda _s}{16\pi
	^2} \right)^2\left( \frac{11}{2}-\sqrt{3}\pi  \right)\approx-6\times
	10^{-5}\,\lambda _s^2,\nonumber \\
	\gamma _2 &=& -\gamma _w-3\gamma _H=-24\left( \frac{\lambda _s}{16\pi
	^2} \right)^2\left( \frac{37}{2}-3\sqrt{3}\pi  \right)\approx-2\times
	10^{-3}\,\lambda _s^2,\nonumber \\
	\gamma _3 &=& -4\gamma _w=24\left( \frac{\lambda _s}{16\pi ^2}
	\right)^2\approx1\times 10^{-3}\,\lambda _s^2,\nonumber \\
	\gamma _4 &=& -2\gamma _w-2\gamma _H=-24\left( \frac{\lambda _s}{16\pi
	^2} \right)^2\left( 12-2\sqrt{3}\pi  \right)\approx-1\times
	10^{-3}\,\lambda _s^2.
\end{eqnarray}
The relative weights with which $\gamma _w$ and $\gamma _H$ appear in the
effective anomalous dimensions $\gamma _1$ and $\gamma _2$ can be read
off from the probabilities with which the normalized states
$w^+w^-,zz/\sqrt{2}$, and $HH/\sqrt{2}$ appear in $\chi _1$ and $\chi
_2$, Eq.\ (\ref{iii45}). The parameter $\sigma $ is defined in Eq.\
(\ref{ii20}). The $a_1$ amplitude arises primarily from the SO(4) singlet
state, and is consistently about three times larger than the SO(4) nonet
amplitudes. We will use these amplitudes in the analysis in the following
section.

\subsection{Argand diagram analysis}

The diagonalized $2\rightarrow2$ scattering amplitudes $a_1$ and $a_3$ from
Eq.\ (\ref{iii50}) are plotted in Figs.\ 3a and 3b as functions of
$\lambda_s$. The anomalous dimension factors
$\sigma ^{\gamma _i}$ have been omitted as they are close to unity for
$\sqrt{s}<100$ TeV and the values of $\lambda_s$ which will be of interest.
The corresponding curves for $a_2$ and $a_4$ are quite close
to that for $a_3$, Fig.\ 3b, and are not shown.

It is clear from Fig.\ 3 that $a_1$ and $a_3$ move away from the
unitarity circle rather quickly as $\lambda_s$ is increased. The contributions
of the $2\rightarrow3$ processes to the unitarity sum in Eq.\ (\ref{ii32})
are quite
small,
\begin{equation}
\sum({\bf a}_0^{2\rightarrow3})^+{\bf a}_0^{2\rightarrow3}=
\left\{
\begin{array}{l}
\frac{\lambda_s}{16\pi^2}(\pi\sqrt{3}-\frac{9}{2})|a_1|^2
\approx 8.5\times10^{-5}\,\lambda_s^3\\
\frac{2\lambda_s}{16\pi^2}|a_3|^2\approx2.0\times10^{-5}\,\lambda_s^3
\end{array}
\right.\label{insert52}
\end{equation}
for scattering from the initial states $\chi'_1$ and $\chi_3$. The
contributions
from $2\rightarrow4$ processes are similarly small \cite{djl2}, less than
$8.0\times10^{-5}\,\lambda_s^4$ and $3.2\times10^{-5}\,\lambda_s^4$ for the
$\chi_1$ and $\chi_3$ channels for $\sqrt{s}<100$ TeV. As a result, the exact
$2\rightarrow2$ amplitudes $a_1$ and $a_3$ must lie essentially on the
unitarity circle for $\lambda_s$ not too large \cite{deltaz}. The deviations of
the calculated amplitudes from the circle give a measure of the range of
$\lambda_s$ in which the calculated amplitudes are reliable.

The convergence---or lack of convergence---of the perturbation series is
illustrated for the channel $\chi' _1\rightarrow\chi _1'$
in the vector diagram for $a_1$ in Fig.\ 4. In this figure,
we show the
zero-, one-, and two-loop amplitudes for $\lambda_s=2.5$ as vectors
in the Argand diagram. The complete two-loop amplitude
$a^{(0)}+a^{(1)}+a^{(2)}$
is the sum of these vectors. It must lie essentially on the unitarity circle if
the perturbative approximation is to be valid. It is immediately evident from
the figure that the series is not converging well for $\lambda_s=2.5$:
$|a^{(2)}|$ is nearly as large as $|a^{(1)}|$ which is as large as
$|a^{(0)}+a^{(1)}|$. Furthermore, the imaginary part of the amplitude becomes
negative for $\lambda_s\agt2.6$. It must be positve in the exact result.

We can quantify the incipient breakdown of the perturbation series
using several tests discussed in \cite{djl2}. We will limit our
attention here to familiar ratio tests which quantify the
observations above about the vector diagram in Fig.\ 4 and its
analogs for the other channels. Further less-general but sometimes
more restrictive tests are discussed in \cite{maher}.

In Figs.\ 5a and 5b, we show the ratios
$|a_i^{(2)}/(a_i^{(0)}+a_i^{(1)})|$ and $|a_i^{(2)}/a_i^{(1)}|$ for the four
amplitudes $a_i$.
For values of $\lambda_s$ greater than 2.3 to 3.2 (depending on the ratio and
channel considered),
the ratios exceed unity and there is no evidence of
convergence of the series from either the ratio of the two-loop amplitude to
the previous partial sum, or from the ratio of successive terms in the
series.
The strongest limits, $\lambda_s<2.3$--2.4 for ratios less
than unity, come from the $\chi_3$ channel. This channel is associated with the
{\bf 5} representation of SO(3), and does not involve $HH$ scattering. We
also note
that Im $a_3$ is negative for $\lambda_s>2.26$.

We will adopt the value $\lambda_{s,\rm max}=2.3$ as the maximum value of
$\lambda_s$ for which the perturbation series in $\lambda_s$ may reasonably be
said to converge at two loops for energies $\sqrt{s}<100$ TeV. At higher
energies, the anomalous-dimension factors in Eq.\ (\ref{iii50})
begin to affect the
magnitudes of the $a_i$.
These factors can be treated in two ways. If the
expression for $\cal M$ is used in the form in Eq.\ (\ref{ii22})
given
by the renormalization group, the anomalous-dimension factors $\Gamma $
multiply each of $a^{(0)}$,
$a^{(1)}$, $a^{(2)}$ and divide out in the ratios considered above, exactly
for $a_3$ and $a_4$, and approximately for the average factors in $a_1$ and
$a_2$. In this approach, the limits above are unchanged. An alternative
procedure is to expand the anomalous dimension factors. This adds an
energy-dependent term to
$a^{(2)}$. In this case the ratio bounds from $a_3$ and $a_4$ are strengthened
and those from $a_1$ and $a_2$ are weakened. For example,
$|a_3^{(2)}/a_3^{(1)}|$ now reaches unity for $\lambda_s=1.85$ for
$\sqrt{s}=10^{16}$ GeV and $M_H=155$ GeV. The change from $\lambda_s=2.3$
becomes irrelevant when $M_H$ is determined by inverting Eq.\ (\ref{ii16})
at the
high energy scale. A general limit $\lambda_{s,\rm max}=2.3$ therefore seems
reasonable.

\subsection{Conclusions}

The restriction $\lambda_s<\lambda_{s,\rm max}=2.3$ is a quantitative
condition on the running coupling which must be satisfied if low-order
perturbation theory in $\lambda_s$ is to give a good description of high-energy
$w^\pm,z,H$ scattering. As illustrated above, the perturbation series for the
$a_i$ show little or no sign of converging at two loops for
$\lambda_s>\lambda_{s,\rm max}$. This of course does not preclude at least
asymptotic convergence for large $\lambda_s$. It is simply that perturbation
theory breaks down as a useful tool, and the theory becomes in that sense
strongly interacting even though the scattering amplitudes are quite small.
Nonperturbative methods are then needed to investigate the problem. However, it
is quite
useful to bound the region in which low-order perturbation theory can be
trusted since that is the method used for most phenomenological calculations.

The limiting value $\lambda_{s,\rm max}=2.3$ for the running coupling in a
weakly interacting or perturbative theory translates directly into an upper
bound on the mass of the Higgs boson
in such a theory once it is decided at what energy the
bound is to be applied. The running coupling $\lambda_s=\lambda_s(s,M_H^2)$ is
plotted in Fig.\ 6 as a function of $M_H$ for various values of
$\sqrt{s}$. Different choices for the upper bound on the perturbative region
correspond to different horizontal lines on this plot. The line for
$\lambda_{s,\rm max}=2.3$ is shown.

Since there is no experimental evidence at the present time for any breakdown
of the standard model, it is plausible---and is generally assumed---that it
will remain valid up to at least a few TeV. If it remains valid up to a
(minimal) value $\sqrt{s}=5$ TeV and we apply the bound $\lambda_s<2.3$ there,
Fig.\ 6 shows that the
Higgs sector of the standard
model will be approximately unitary and weakly
interacting at two loops only for $M_H<380$ GeV. This is signifcantly below the
values $M_H \approx 1$ TeV used in a number of phenomenological studies done
using tree-level amplitudes. These results should be re\,examined. The limit is
also below the nonperturbative ``triviality" bound $M_H\alt650$--800 GeV
obtained in \cite{hasen}. The existence of a Higgs boson with a mass in the
nonperturbative or strongly interacting sector between our bound and the
triviality limit is not precluded, and would be quite interesting. Calculations
would be difficult.

The bound on $M_H$ in a perturbative theory becomes much stronger,
$M_H\alt155$ GeV, if the standard model is assumed to hold up to a
typical unification energy of order $10^{16}$ GeV. This result may be
altered slightly when other couplings are included in the renormalization
group equations, but the Higgs boson will still remain light.

Finally, as noted elsewhere \cite{djl2,ldjl}, the limit on the range of
validity of
perturbation theory can become a real upper bound on $M_H$ if the standard
model is embedded in a specific unified or dynamical model, which
{\em must\/} remain perturbative up to some unification or dynamical energy. If
this is the actual situation, the Higgs boson cannot be much more massive than
380 GeV
in theories such as technicolor with a low mass scale, and will be much
lighter in typical unified theories. It
will then be accessible in future experiments.

\acknowledgements

This work was supported in part by the U.S. Department of Energy under Contract
No.\ AC02-76ER00881. One of the authors (L.D.) would like to thank the Aspen
Center for Physics for its hospitality while parts of this work were done.

\appendix
\section*{CALCULATION OF \mbox{\lowercase{$\gamma _w$}},\ $\gamma _H$, and
$\beta$ AT TWO LOOPS}\label{appendixa}

Because of our on-mass-shell convention for the renormalization of the
scalar theory, we will determine $\beta $ and the $\gamma $'s to two
loops using the Callan-Symanzik definitions for these quantities
\cite{chli},
%
% eqnarray
%
\begin{eqnarray}
\label{a1}
	\beta  &=& \lim_{\epsilon \rightarrow0}\left(
	2M_H^2\frac{\partial\lambda /\partial M_0^2}{\partial M^2/\partial
	M_0^2} \right)_{\lambda _{0,\epsilon }},\\
	\gamma _i &=& \lim_{\epsilon \rightarrow0}\left( M_H^2\frac{\partial
	Z_i/\partial M_0^2}{\partial M_H^2/\partial M_0^2} \right)_{\lambda
	_{0,\epsilon }}.
\end{eqnarray}
Here $M_H$ and $\lambda $ are the renormalized or physical Higgs boson
mass and coupling. $M_0$ and $\lambda _0$ are the corresponding bare
quantities. Since the Goldstone boson mass $m_w$ is fixed at zero, $M_H$
is the only mass in the physical theory.

The bare coupling $\lambda _0$ is given in \cite{mdr} as
%
% eqnarray
%
\begin{eqnarray}
\label{a3}
	\lambda _0 &=& \lambda +\frac{\lambda ^2\xi ^\epsilon }{16\pi ^2}\left(
	\frac{12}{\epsilon }+25-12\gamma -3\sqrt{3}\pi +\epsilon [\,3\pi
	\sqrt{3}\ln3-12\sqrt{3}{\bf C}-6\pi \sqrt{3}(1-\case{1}/{2}
	\gamma ) \right.\nonumber \\
	&&\left.
	-\pi ^2+6\gamma ^2-25\gamma +\case{99}/{2}]+{\rm O}(\epsilon ^2)\right)
	+\frac{\lambda ^3\xi ^{2\epsilon }}{(16\pi ^2)^2}\left(
	\frac{144}{\epsilon ^2}+\frac{18}{\epsilon }(29-16\gamma -4\sqrt{3}\pi )
	 \right.\nonumber \\
	 && \frac{32906}{27}+12K_3+162K_5+36K_2-90\zeta (3)\nonumber \\
	 &&+\frac{185}{3}\zeta (2)-1044\gamma +288\gamma ^2+54\pi
	 ^2+\frac{1184}{9}\ln2\nonumber \\
	 &&+72\ln^2\left( \frac{\sqrt{5}+1}{2} \right)-114\sqrt{5}\ln\left(
	 \frac{\sqrt{5}+1}{2} \right)\nonumber \\
	 &&+81\zeta (2)\ln2-369\pi \sqrt{3}+144\pi \sqrt{3}\gamma
	 -60\sqrt{3}{\bf C}\nonumber \\
	 &&\left.+72\pi \sqrt{3}\ln3+{\rm O}(\epsilon )\right)+{\rm O}(\lambda ^4)
\end{eqnarray}
to two loops. Here $\xi =4\pi \mu ^2/M_H^2$, and $\mu $ is the mass scale
introduced to keep the coupling $\lambda $ dimensionless in
$d=4-2\epsilon $ dimensions. The renormalization constants $Z_w$ and
$Z_H$ are \cite{mdr}
%
% eqnarray
%
\begin{eqnarray}
\label{a4}
	Z_w &=& 1-\frac{\lambda }{16\pi ^2}\xi ^\epsilon
	\left[ 1+\epsilon \left(
	\case{3}/{2}-\gamma  \right)+{\rm O}(\epsilon ^2) \right]\nonumber \\
	&&+\frac{\lambda ^2}{(16\pi ^2)^2}\xi ^{2\epsilon }\left( -\frac{3}{\epsilon
	}+42\sqrt{5}\ln\left( \frac{\sqrt{5}+1}{2} \right)-96\ln^2\left(
	\frac{\sqrt{5}+1}{2} \right) \right.\nonumber \\
	&&\left.+\frac{400}{9}\ln2+6\gamma -\frac{38}{3}\zeta
	(2)-12K_3-\frac{1525}{54}+3\sqrt{3}\pi +{\rm O}(\epsilon )\right),
\end{eqnarray}
and
%
% eqnarray
%
\begin{eqnarray}
\label{a5}
	Z_H &=& 1+\frac{\lambda }{16\pi ^2}\xi ^\epsilon \left[ 12-2\pi
	\sqrt{3}\right.\nonumber \\
	&&\left.+\epsilon (24-12\gamma -3\pi \sqrt{3}(1-\case{2}/{3}\gamma
	)+2\pi \sqrt{3}\ln3-8\sqrt{3}{\bf C})+{\rm O}(\epsilon ^2) \right]
	\nonumber \\
	&&+\frac{\lambda ^2}{(16\pi ^2)^2}\xi ^{2\epsilon }\left(
	-\frac{3}{\epsilon }+144\ln2+42\pi ^2+81\zeta (2)\ln2 \right.\nonumber \\
	&&+36K_2+162K_5+6\gamma +33\zeta (2)-90\zeta (3)\nonumber \\
	&&\left.-16\sqrt{3}\pi \ln3+334\sqrt{3}{\bf C}+\frac{273}{2}-292\sqrt{3}\pi
	+{\rm O}(\epsilon )\right).
\end{eqnarray}
Finally, $M_0$, which we will not need explicitly, is given in terms of
the self-energy functions $\Pi _w^0$ and $\Pi _H^0$ for the bare fields
calculated in \cite{mdr} by \cite{mass}
%
% equation
%
\begin{equation}
\label{a6}
	M_0^2=M_H^2-{\rm Re}\Pi _H^0(M_H^2)+\Pi _w^0(0).
\end{equation}

We have calculated $\beta $ and $\gamma $'s using two different
approaches. The first is based on the expressions for the $Z$'s and
$\lambda _0$ above. These give the first few terms in series of the form
%
% eqnarray
%
\begin{eqnarray}
\label{a7}
	\lambda _0 &=& \mu ^{2\epsilon }\lambda \left(
	1+\sum_{j=1}^{\infty}\sum_{k=0}^{j}a_{jk}\frac{\left( \lambda \xi
	^{\epsilon } \right)^j}{\epsilon ^k} \right),\\
	Z &=& 1+\sum_{j=1}^{\infty}\sum_{k=0}^{j-1}c_{jk}\frac{\left( \lambda
	\xi ^\epsilon  \right)^j}{\epsilon ^k}.\label{a8}
\end{eqnarray}
$M_H$ and $\lambda $ are regarded in Eqs.\ (\ref{a1}) and (A2) as
implicit functions of the independent variables $M_0$ and $\lambda _0$
\cite{chli}; $\mu $ and $\epsilon $ are fixed. Differentiation of the
equation for $\lambda _0$ gives
%
% equation
%
\begin{equation}
\label{a9}
	\frac{d\lambda _0}{dM_0^2}=0=\frac{\partial \lambda _0}{\partial \lambda
	}\,\frac{\partial \lambda }{\partial M_0^2}-\frac{\xi
	}{M_H^2}\,\frac{\partial \lambda _0}{\partial \xi }\,\frac{\partial
	M_H^2}{\partial M_0^2},
\end{equation}
or from Eq.\ (\ref{a1}),
%
% equation
%
\begin{equation}
\label{a10}
	\beta =\lim_{\epsilon \rightarrow0}2\xi \frac{\partial \lambda
	_0/\partial \xi }{\partial \lambda _0/\partial \lambda },
\end{equation}
where $\lambda _0$ is defined in the last expressions by the series in
Eq.\ (\ref{a7}). Substituting the series into Eq.\ (\ref{a10}), expanding
to order $\lambda ^3$, and taking the limit $\epsilon \rightarrow0$ gives
%
% eqnarray
%
\begin{eqnarray}
\label{a11}
	\beta  &=&
	\lambda ^2(2a_{1,1}+4\lambda a_{2,1}-8\lambda a_{1,1}a_{1,0})\nonumber \\
	 &=& 24\frac{\lambda ^2}{16\pi ^2}\left( 1-
	 13\frac{\lambda }{16\pi ^2} \right)\nonumber \\
	 &\equiv& \beta _0\lambda ^2+\beta _1\lambda ^3.
\end{eqnarray}
A potentially singular term of the form
%
% equation
%
\begin{equation}
\label{a12}
	\frac{\lambda ^2\xi ^{2\epsilon }}{\epsilon }\left( a_{2,2}-a_{1,1}^2 \right)
\end{equation}
vanishes identically in this construction because $a_{2,2}\equiv
a_{1,1}^2$. The result is therefore finite and independent of the
arbitrary mass $\mu $, as expected.

A similar calculation gives
%
% eqnarray
%
\begin{eqnarray}
\label{a13}
	\gamma _i &=& \lim_{\epsilon \rightarrow0}\left( \xi \frac{\partial
	Z_i}{\partial \xi }+\xi \frac{\partial \lambda _0/\partial \xi
	}{\partial \lambda _0/\partial \lambda }\frac{\partial Z}{\partial \lambda }
\right)\nonumber \\
	 &=& \lambda ^2(a_{1,1}c_{1,0}-2c_{2,1}),
\end{eqnarray}
hence,
%
% eqnarray
%
\begin{eqnarray}
\label{a14}
	\gamma _w &=& -6\left( \frac{\lambda }{16\pi ^2} \right)^2\\
	\gamma _H &=& (150-24\pi \sqrt{3})\left( \frac{\lambda }{16\pi ^2}
	\right)^2.\label{a15}
\end{eqnarray}

We note finally that the mass dimension $\gamma _M$, defined as
%
% equation
%
\begin{equation}
\label{a16}
	\gamma _M=1-\frac{M_H^2/M_0^2}{\partial M_H^2/\partial M_0^2},
\end{equation}
is connected to $\beta $ and $\gamma _w$ through the relation $\lambda
_0=\left( M_0^2/M_H^2Z_w \right)\lambda $. This gives
%
% eqnarray
%
\begin{eqnarray}
\label{a17}
	\gamma _M &=& \beta /2\lambda -\gamma _w\nonumber \\
	 &=& 12\frac{\lambda }{16\pi ^2}-150\left( \frac{\lambda }{16\pi ^2}
	 \right)^2.
\end{eqnarray}

A second approach to the calculations above uses the observation that the
introduction of the mass $\mu $ in the course of dimensional
regularization is unnecessary. With our on-shell renormalization scheme,
the theory contains a single mass $M_H$ which can be used to scale the
coupling in the transition to $4-2\epsilon $ dimensions. With the choice
$\mu ^2=M_H^2/4\pi $, the quantity $\xi $ disappears from the theory, and
%
% eqnarray
%
\begin{eqnarray}
\label{a18}
	\lambda _0 &=& \lambda \left( \frac{M_H^2}{4\pi } \right)^\epsilon
	\left( 1+\sum_{j=1}^{\infty}\sum_{k=0}^{j}a_{jk}\frac{\lambda
	^j}{\epsilon ^k} \right).\\
	Z &=& 1+\sum_{j=1}^{\infty}\sum_{k=0}^{j-1}c_{jk}\frac{\lambda
	^j}{\epsilon ^k}.
\end{eqnarray}
Using the Callan-Symanzik definitions, we find that
%
% equation
%
\begin{equation}
\label{a20}
	\beta =\lim_{\epsilon \rightarrow0}\,\beta (\epsilon
	)\equiv\lim_{\epsilon \rightarrow0}\frac{-2\epsilon \lambda _0}{\partial
	\lambda _0/\partial \lambda }
\end{equation}
and
%
% equation
%
\begin{equation}
\label{a21}
	\gamma _i=\lim_{\epsilon \rightarrow0}\,\frac{1}{2}\beta (\epsilon
	)\frac{\partial Z_i}{\partial \lambda }.
\end{equation}
It is straightforward to show that the limits of these expressions are
identical to all orders in $\lambda $ to those obtained from Eqs.\
(\ref{a10}) and (\ref{a13}). They are also formally identical to the
expressions obtained in mass-independent subtraction schemes with
dimensional regularization \cite{nomass}, and imply in the same way
recurrence relations for the coefficients of powers of $1/\epsilon $
greater than the first \cite{chli}.

\begin{figure}
	\caption{The three distinct topologies for the tree-level $2\rightarrow3$
processes. The exchange graphs (a) and (b) are suppressed in the
$2\rightarrow3$ cross section or unitarity sum for $s\rightarrow\infty$.
The ``jet" graph (c) is not suppressed because of the contributions of
low-mass pairs to the jet.}
\end{figure}

\begin{figure}
\caption{Inelastic $2\rightarrow3$ diagrams with the topology of cut
{\em Eye\/} graphs for initial $zH$ states. The terminology is that of Maher
{\em et al.\/} [18]. We show:
%\cite{mdr}.
(a), the cut ${\cal E}_2$ graph which contributes to
$-Z_w^{(1)}|a_{zH}^{(0)}|^2$; (b), (c), the cut ${\cal E}_4$ and ${\cal E}_2^*$
graphs which contribute to $-Z_H^{(1)}|a_{zH}^{(0)}|^2$.}
\end{figure}

\begin{figure}
	\caption{Behavior of the $j=0$ eigenamplitudes  for
$w^\pm,z,H$ scattering shown as functions of the running coupling
$\lambda _s=\lambda _s\left( s,M_H^2 \right)$: (a) $a_1$ and (b) $a_3$.
The factors associated
with the anomalous dimensions are very close to unity for
$\protect\sqrt{s}$ in the TeV
range and have been neglected. The curves for $a_2$ and $a_4$ are very
similar to that for $a_3$ shown in (b).}
\end{figure}

\begin{figure}
	\caption{Vector diagram showing the decomposition of the eigenamplitude
$a_1$ into tree-level, one-loop, and two-loop contributions
$a^{(0)},a^{(1)}$, and $a^{(2)}$. The dashed curve is a segment of the
unitarity circle. Corresponding values of the running couplings $\lambda
_s$ are shown along the real axis for $a^{(0)}$, along the upper solid
curve for $a^{(0)}+a^{(1)}$, and along the lower solid curve for
$a^{(0)}+a^{(1)}+a^{(2)}$.}
\end{figure}

\begin{figure}
	\caption{(a) The ratio of the magnitude $|a^{(2)}|$ of the two-loop
contribution to the $j=0$ scattering amplitude
$a_i=a^{(0)}+a^{(1)}+a^{(2)}+\cdots$ to the magnitude $|a^{(0)}+a^{(1)}|$
of the complete amplitude at one loop, shown as a function of the running
coupling $\lambda _s$. (b) The ratio $|a^{(2)}|/|a^{(1)}|$ of the
magnitudes of the two- and one-loop contributions to the scattering
amplitude as a function of $\lambda _s$. Requiring that each term
of the perturbation series be smaller than the previous term (or
previous partial sum), gives a limit $\lambda _s\protect\alt2.3$ for the
$\chi _3\rightarrow\chi _3$ channel.}
\end{figure}

\begin{figure}
	\caption{The relation between the running coupling $\lambda _s$ and the
physical mass of the Higgs boson $M_H$ for various values of the energy
at which the unitarity constraint could be applied. The horizontal line is at
$\lambda _s=2.3$ which we adopt as the maximum value of $\lambda
_s$ for which the perturbation series in $\lambda _s$ may
reasonably be said to converge.}
\end{figure}


\begin{references}
\bibitem{algebra}
The rather complex algebraic manipulations involved in the calculations
reported here were performed using the computer program REDUCE.
\bibitem{nomass}
For mass-independent subtraction schemes, $\lambda_0$ and $Z$ are
given by series identical in form to those in Eqs.\ (A18) and (A19)
with $M_H^2/4\pi$ replaced
in Eq.\ (A18)
by the scale parameter $\mu^2$, while $\beta$
and $\gamma$ are defined by
\[
\beta = \lim_{\epsilon\rightarrow0} \mu
\frac{\partial\lambda}{\partial\mu}\quad,
\quad
\gamma = \lim_{\epsilon\rightarrow0} \case{1}/{2}\mu
\frac{\partial Z}{\partial\mu}.
\]
Use of the equation $\partial\lambda_0/\partial\mu = 0$ gives
expressions identical to those in Eqs.\ (A20) and (A21). The only
difference between the on-mass-shell and mass-independent
renormalization schemes is in the specific values of the
coefficients $a_{jk}$ and $c_{jk}$.
\end{references}
\end{document}